\documentclass[a4paper]{article}

\usepackage{amsmath,amssymb}

\newcommand{\ii}{\mathrm{i}}

\newcommand{\dd}{\mathrm{d}}
\newcommand{\pd}{\partial}
\newcommand{\hh}{\mathcal{H}}

\newcommand{\e}{\mathrm{e}}
\newcommand{\ket}[1]{\left|#1\right\rangle}

\newcommand{\tr}{\mathop{\mathrm{tr}}\nolimits}

\newcommand{\q}{\tilde{\theta}}
\newcommand{\sgn}{\mathop{\mathrm{sgn}}}
\newcommand{\drc}{\hat{\partial}}
\begin{document}

\title{Fermion Doubling and Berenstein--Maldacena--Nastase Correspondence}
\author{Stefano Bellucci\thanks{e-mail:
\texttt{bellucci@lnf.infn.it}}, and Corneliu Sochichiu\thanks{On
leave from: \textit{Bogoliubov Laboratory of Theoretical Physics,
Joint Institute for Nuclear Research, 141980 Dubna, Moscow Reg.,
RUSSIA}; e-mail: \texttt{sochichi@lnf.infn.it}}
\\
\\
{\it INFN-Laboratori Nazionali di Frascati,}\\
{\it Via E. Fermi 40, 00044 Frascati, Italy}}

\maketitle
\begin{abstract}
We show that the string bit model suffers from doubling in the
fermionic sector. The doubling leads to strong violation of
supersymmetry in the limit $N\to\infty$. Since there is an exact
correspondence between string bits and the algebra of BMN
operators even at finite $N$, doubling is expected also on the
side of super-Yang--Mills theory. We discuss the origin of the
doubling in the BMN sector.
\end{abstract}

\section{Introduction}

Large $N$ physics \cite{'tHooft:1974jz,'tHooft:2002yn} plays an
important role in the correspondence between Yang--Mills theory
and strings (see \cite{Aharony:1999ti} for a review).
Recently, Berenstein--Maldacena--Nastase (BMN) advocated in
\cite{Berenstein:2002jq,Berenstein:2002sa,Berenstein:2002zw} that
IIB superstring theory on pp-wave background can be described in
terms of a particular set of operators (BMN operators) having
large $R$-charge $J$ in super--Yang--Mills theory. The pp-wave
background \cite{Blau:2001ne,Blau:2002dy,Blau:2002mw}, appears as
a Penrose limit of anti-de Sitter (AdS) space. Therefore, the BMN
correspondence can be seen as a limit of the AdS/CFT correspondence
\cite{Maldacena:1998re,Gubser:1998bc}. The peculiarity of this
background is that string theory can be solved there
\cite{Metsaev:2001bj,Metsaev:2002re}.

The BMN correspondence was conjectured to hold in the limit of
large (infinite) $J$ and $N$, the quantity $g_{BMN}=J^2/N$ being
the effective coupling of the string interaction. For finite
values of $J$ and $N$, however, the dynamics of BMN operators was
shown to be equivalent to the string bit model
\cite{Thorn:1997jy,Verlinde:2002ig,Zhou:2002mi,Vaman:2002ka,Pearson:2002zs}.
The string bit model was introduced by Thorn \cite{Thorn:1997jy},
as a supersymmetric mechanical model describing fragmented
superstring. For earlier works dealing with the discretization of one
dimensional superspace, see \cite{B23,B22,B21}.

On the other hand, it is not difficult to see that the above
string fragmentation corresponds to the lattice discretization of
the string. In the Green--Schwarz approach, IIB superstring
contains, besides the bosonic fields, also fermionic ones in target space
representation.\footnote{For a useful parametrization of the scalar
superfields involved, see \cite{B1}.}
It is well known, however, that the lattice
formulation of systems that include fermions possesses a strong
drawback, related to a fermion doubling problem (see e.g.
\cite{Montvay:lat}). In particular, the lattice formulation of
\emph{supersymmetric theories} faces the problem of fermion
doubling (see \cite{Montvay:2001aj,Montvay:1998ak} and references
therein for a review of the problem and subsequent developments).
Thus, one may expect that the string bit model is spoiled by
fermion doubling too. Below, we show that this is indeed the case.

Having in mind the abovementioned fact, together with the
assumption that the string bit model describes \emph{exactly} the
BMN operator dynamics, we can conjecture that there are wrong
fermionic modes in the BMN sector of super--Yang--Mills theory,
which survive in the large $N$ (or large $J$) limit. This
conjecture is supported also by the fact that computations in the
BMN approach are essentially the same as in matrix theory, while
at the same time in the latter one can find traces of the
fermionic doubling \cite{Sochichiu:2000fs}.

The plan of the paper is as follows. Firstly, we briefly review
the string bit model in both Hamiltonian and Lagrangian
approaches. In the third section we analyze the fermionic spectrum
and discover the low energy fermionic states corresponding to
large lattice momenta (the edge of the Brillouin zone). Next, we
solve the equations of motion for the bits, which allows one to
quantize the model the same way, as it was done in the case of the
continuous string. In section four we discuss the fate of
supersymmetry in the case of the fermionic doubling. Using a
simplified version of sting bits as a toy-model, we show that the
contribution of the fermionic mirror states leads to a strong violation
of supersymmetry, in the continuum limit. After that, we show how
doubling states can appear in the BMN correspondence and, finally,
we discuss the results.
\section{Bit String Model}
Let us shortly review the pp-wave IIB superstring bit model
\cite{Verlinde:2002ig}. The superstring consisting of $J$ bits is
described in terms of phase space coordinates  of its bits and
their superpartners: $\{p^i_n,x^i_n, \theta^a_n, \q^a_n\}$, where
$n=0,\dots,J-1$. The phase space variables satisfy the following
(classical) commutation relations:
\begin{equation}\label{PB}
  [p^i_n,x^i_n]=\delta^{ij}\delta_{mn},\quad
  \{\theta^\alpha_n,\theta^\beta_m\}=\frac{\ii}{2}\delta^{ab}\delta_{mn},\quad
  \{\q^a_n,\q^b_m\}=\frac{\ii}{2}\delta^{ab}\delta_{mn}.
\end{equation}

The reparametrization invariance of the string becomes, in the bit
language, the invariance with respect to the symmetry group $S_J$
of permutations of the labels $n$. The whole permutation group
$S_J$ is split into equivalence classes $[\gamma]$ of permutations
having the same number of cycles with fixed lengths
$J_1,J_2,\dots, J_s$, $\sum J_k=J$. Then, the Hilbert space of the
quantized model can be split, according to this, into a direct sum
of \emph{twisted sectors} $\hh_\gamma$, associated with each
conjugacy class $[\gamma]$. The transformations inside a conjugacy
class reduces to relabelling of bits in the cycles. Therefore, the
twisted sector corresponding to the conjugacy class $[\gamma_s]$
with $s$ cycles, can be identified with the $s$-string Hilbert
space. In particular, one string sector corresponds to cyclic
subgroups of $S_J$, and up to relabelling of $n$ is given by
\begin{equation}\label{gamma}
  \gamma_1(n)=n+1 \mod J.
\end{equation}

In the case of the $s$-string sector, one can introduce the
following standard $\gamma_s$ transformation by fixing the
representant of the conjugacy class $[\gamma_1]$. For this, let us
relabel $0\leq n \leq J-1$ by sets of bits $\{n_k;$ $0\leq n_k
\leq J_k-1$, $k=1,\dots,s\}$ and define
\begin{equation}\label{multi-s}
  \gamma_s=\gamma_1^{(1)}\gamma_1^{(2)}\dots\gamma_1^{(s)},\quad
  \gamma_1^{(k)}(n_k)=n_k+1 \mod J_k.
\end{equation}
Since the conjugation transformations preserve the cyclic
structure of $\gamma_s$ (including the lengths of the cycles), just
changing the labels \cite{Scott:GT}, in order to get the arbitrary
representative $\gamma'_s$ of the conjugacy class, it suffices to
replace each bit label by a permutation: $n\mapsto\sigma(n)$.
Among the permutations, however, there are some which do not
change the cycles, thus leaving us with the same $\gamma_s$.

The Hamiltonian and supercharges describing the model read as
follows: \cite{Verlinde:2002ig},
\begin{subequations}\label{h-and-su}
\begin{align}\label{h}
 &H=H_B+H_F,\\
 \label{Q}
 &Q=\sum_{n=0}^{J-1}[a(p^i_n\gamma_i\theta_n-x^i_n\gamma_i\Pi\q_n)+
 (x^i_{\gamma(n)}-x^i_n)\gamma_i\theta_n],\\
 &\tilde{Q}=\sum_{n=0}^{J-1}[a(p^i_n\gamma_i\q_n-x^i_n\gamma_i\Pi\theta_n)-
 (x^i_{\gamma(n)}-x^i_n)\gamma_i\q_n],
\end{align}
\end{subequations}
where
\begin{align}\label{hb}
  &H_B=\sum_{n=0}^{J-1}[\frac{a}{2}(p_{in}^2+x_{in}^2)+\frac{1}{2a}(x^i_{\gamma(n)}-x^i_n)^2],\\
  \label{hf}
  &H_F=-\ii \sum_{n=0}^{J-1}[(\theta_n\theta_{\gamma(n)}-
  \q_n\q_{\gamma(n)})-2a\q_n\Pi\theta_n].
\end{align}

In the Lagrangian form the action compatible with the Hamiltonian
\eqref{h} and the commutation relations \eqref{PB} are given by
\begin{multline}\label{action}
  S=\sum_{n=0}^{J-1}\left[\frac{a}{2}\dot{x}^2_{in}-
  \frac{1}{2a}(x^i_{\gamma(n)}-x^i_n)^2-\frac{a}{2}x^2_{in}\right.\\
  +\left.\ii a(\theta_n\dot{\theta}_n+\q_n\dot{\q}_n)+
  \ii(\theta_n\theta_{\gamma(n)}-\q_n\q_{\gamma(n)})+2\ii a\q_n\Pi\theta_n\right].
\end{multline}
The expressions \eqref{h-and-su} correspond to a discrete version
of IIB superstring in pp-wave background
\cite{Metsaev:2001bj,Metsaev:2002re}, obtained by the most
straightforward (naive) discretization. Since this naively
discretized model contains fermions, one should expect problems
typical of lattice fermions.

\section{Fermion doubling}

Let us analyze the fermionic spectrum in the free string bit
model. In order to do this, let us consider the fermionic part
\eqref{hf} of the Hamiltonian and ``diagonalize'' it. For this,
let us perform the bit (lattice) Fourier transform of the fields
\begin{subequations}\label{fourier}
\begin{align}
  &\theta_n=\frac{1}{\sqrt{J}}\sum_{p=-J/2}^{J/2} \theta_p\e^{2\pi\ii ln/J},\\
  &\q_n=\frac{1}{\sqrt{J}}\sum_{p=-J/2}^{J/2} \q_p\e^{2\pi\ii pn/J},
\end{align}
\end{subequations}
where, by abuse of notations, we kept the same character for both
the field and its Fourier transform, distinguishing them only by
the labels
\begin{align}\label{x-rep}
  &l,m,n,\dots=0,1,\dots,J-1;\quad &\text{($x$-representation)},\\
  \label{p-rep}
  &p,q,r,\dots=-J/2,-J/2+1,\dots,J/2;\quad
  &\text{($p$-representation)}.
\end{align}
From \eqref{p-rep} one can notice that for odd $J$ the ``momenta''
$p,q,r,\dots$ run through integer numbers, while for an even $J$ value,
they should be half-integer. This has no particular meaning and is
a result of the choice for the origin of the momentum space,
which in the present case was taken to be symmetric with respect
to the inversion of momenta $p\to -p$.

Let us consider, for definiteness, the one string sector and fix
the standard choice \eqref{gamma} for the ``moduli'' of $\gamma$
permutation. As we discussed above, all other situations in the
same class $[\gamma]$ are obtained from the standard one by all
possible relabelling of bits $n'=n'(n)$. (The other multi-string
sectors can be analyzed in a similar way, by fixing the
``standard'' $\gamma$-permutations to \eqref{multi-s}, and then
``shuffling'' the labels, in order to generalize the result to
arbitrary $\gamma_s$.)

Plugging the transformations \eqref{fourier} in the fermionic
Hamiltonian \eqref{hf}, yields the expression
\begin{multline}\label{hf-four}
  H_F=\sum_p\sin\left(\frac{2\pi
  p}{J}\right)(\theta_{-p}\theta_p-\q_{-p}\q_p)+2a\ii\q_{-p}\Pi\theta_p=\\
  \begin{pmatrix}
    \theta_{-p}& \q_{-p}
  \end{pmatrix}
  \begin{pmatrix}
    \sin\frac{2\pi p}{J}& \ii a\Pi \\
    -\ii a\Pi & -\sin\frac{2\pi p}{J}
  \end{pmatrix}
  \begin{pmatrix}
    \theta_p\\
    \q_p
  \end{pmatrix}.
\end{multline}
It is not difficult to see that the spectrum of the Hamiltonian
\eqref{hf-four} reads
\begin{equation}
  E^{(J)}_p=\pm \sqrt{J^2\sin^2\frac{2\pi p}{J}+1}.
\end{equation}
As expected, in the limit of large $J$, one can expand the sin
function under the square root, in order to get the continuum energy
levels of the fermions\footnote{Notice the difference in notations
with the paper \cite{Metsaev:2001bj}.}
\begin{equation}\label{zero}
  \omega_n=E^{(J)}_{p=n\ll J}\approx \pm\sqrt{(2\pi n)^2+1},
\end{equation}
obtained by Metsaev in \cite{Metsaev:2001bj}.

Eq. \eqref{zero} yields a correct, although incomplete, energy
spectrum for the continuous superstring. Due to the other zero of
the sin function when its argument approaches $\pm\pi$, there are
other low energy levels which survive in the continuum limit
$J\to\infty$. They appear when the momentum $p$ is in the vicinity
of the edge of the \emph{Brillouin zone}, $p\sim \pm J/2$. This
will appear as a 2-fold degeneracy of each energy level in
\eqref{zero}, $E_{p}=E_{J/2-p}$. This phenomenon has been is known
for long time in lattice theories with fermions, where it is
called fermion spectrum \emph{doubling} (for more details see the
textbook \cite{Montvay:lat}).

The doubling can be related to a symmetry of the discrete system
of string bits which relates fermionic modes of different
chiralities \cite{Makeenko:1997bk}
\begin{equation}
  \begin{pmatrix}
    \theta_n \\
    \q_n
  \end{pmatrix}
 \mapsto
 (-1)^{n}
  \begin{pmatrix}
    \Pi \q_n \\
    \Pi \theta_n
  \end{pmatrix}.
\end{equation}
Thus, in the continuum limit, we obtained not just pp-wave IIB
superstring but something more, i.e. the Green--Schwarz superstring
with two fermionic sectors!

In this context one may ask, what happens to supersymmetry? The
short answer is that the lattice theory in fact is not
supersymmetric owing to the effects of discreetness. Also due to
the doubling, the symmetry has few chances to be restored in the
continuum limit!

\section{A note on Supersymmetry and Doubling}

In order to illustrate the behavior of supersymmetry on the
lattice let us consider a simpler toy model example, which catches
however the most important features generic for all supersymmetric
models on the lattice.

Let us consider the model of ``one dimensional superstring''
described by the continuum action
\begin{equation}\label{1ds}
  S=\int\dd^2\sigma\left(\frac{1}{2}\pd_a X\pd_a
  X+\frac{\ii}{2}\psi\drc\psi\right),
\end{equation}
where $X$  and $\psi$ are, respectively, a bosonic field and a
Majorana-Weyl fermion on a two dimensional cylinder. The action
\eqref{1ds} is invariant with respect to the supersymmetry
transformations
\begin{align}\label{susy}
  &\delta X=-\ii\epsilon\psi,\\
  &\delta \psi=\drc X\epsilon,
\end{align}
where $\drc$ is the two dimensional Dirac operator,
$\drc=\gamma_a\pd_a$, and $\epsilon$ is the supersymmetry
transformation parameter, which is a Majorana-Weyl spinor.
In the canonical formalism the system is represented by the
canonical variables $\Pi(\sigma)=(\pd L/\pd \dot{X})$, $X(\sigma)$
and $\psi$, satisfying
\begin{equation}
  [\Pi(\sigma),X(\sigma')]_{PB}=\delta(\sigma-\sigma'),\quad
  \{\psi_\alpha(\sigma),\psi_\beta(\sigma')\}_{PB}=
  \frac{\ii}{2} \gamma^0_{\alpha\beta}\delta(\sigma-\sigma') ,
\end{equation}
and the Hamiltonian
\begin{equation}\label{toyh}
  H=\oint
  \dd\sigma\left(\frac{1}{2}\Pi^2+\frac{1}{2}(X')^2
  +\ii\psi\gamma_1\psi'\right).
\end{equation}

Supersymmetry is generated by the supercharge
\begin{equation}\label{toy-su}
  Q=\oint\dd \sigma[-\ii\Pi\psi+\ii X'\gamma_1\gamma_0\psi],
\end{equation}
which satisfies the (classical) algebra,
\begin{equation}\label{QQ}
  \{Q,Q\}=-2H\gamma_0+2P\gamma_1,
\end{equation}
where $H$ is the Hamiltonian \eqref{toyh} and $P=\Pi X'$ denotes
the shift generator.
Just like the action, the Hamiltonian \eqref{toyh} is invariant
with respect to the supersymmetry transformation
\begin{equation}
  \delta H=\epsilon [Q,H]=0.
\end{equation}

Let us consider now a version of the above model in the case of a
discrete spatial extension $\sigma\equiv\sigma_1$.\footnote{This
model suffers from doubling, in the same way as the string bit
model in the previous section.} In order to do this, let us start
with the supercharge\footnote{We consider the following
discretization of $\sigma$: $\sigma=an$, $n=0,\dots,J$, and
$a=2\pi L/J$.}
\begin{equation}\label{discr-su}
  Q=a\sum_{n=0}^J\left(-\ii
  \Pi_n\psi_n+\frac{\ii}{a}(X_{n+1}-X_n)\gamma_1\gamma_0\psi_n\right).
\end{equation}
This expression is analogous to the supercharge \eqref{Q} and is a
straightforward discretization of \eqref{toy-su}. The discrete
Hamiltonian and the shift operator can be defined through the
lattice version of Eq. \eqref{QQ}. Indeed, for the Hamiltonian one
has
\begin{equation}\label{toy-discr-h}
  H=a\sum_n\left(\frac{1}{2}\Pi^2_n+\frac{1}{2a^2}(X_{n+1}-X_n)^2+
  \frac{\ii}{2a}\psi_n\gamma_1\psi_{n+1}\right),
\end{equation}
while $P$ appears to be the operator of the forward lattice shift,
i.e. $P=\sum_n\Pi_n(X_{n+1}-X_n)$.

The above results agree perfectly with what can be expected from a
naive discretization of the Hamiltonian \eqref{toyh}. However, an
unpleasant surprise comes next. The discrete Hamiltonian
\eqref{toy-discr-h} fails to be exactly supersymmetric! Indeed, a
straightforward computation yields
\begin{multline}\label{noninv}
  \delta H/\delta\epsilon=\\
  \sum_n\left[\frac{\ii}{2}\left(-\Pi_n\gamma_0+
  \frac{1}{a}(X_{n+1}-X_n)\gamma_1\right)
  \gamma_1(\psi_{n+1}-2\psi_n+\psi_{n-1})\right].
\end{multline}

For slowly varying fields (which correspond to smooth functions in
the continuum limit), this part of the supersymmetry variation is
of order $\sim 1/J$ and thus it vanishes, as $J$ approaches
infinity. This occurs because the terms in \eqref{noninv}
correspond to lattice analogs of second derivatives, multiplied by
factors of order $a=2\pi/J$. In the continuum limit they are
supposed to give Lorentz noninvariant terms vanishing like
\begin{equation}
  -\frac{\ii a}{2}\oint\dd \sigma
  [(\Pi\gamma_0-X'\gamma_1)\gamma_1\psi''
  \sim O(1/J).
\end{equation}

This is what would happen, if the doubler states would not come into the
game. For the doubler states the fermionic factor in the r.h.s of
\eqref{noninv} is of the order of unity, while the summation adds a
factor of order $J$ making the non-invariant contribution divergent. This
is in contrast with the situation of the ``genuine'' non-doubled
part, where the fermionic factor is of order $1/J^2$, while
the summation just reduces the decay by one power in $J$.
In conclusion, the supersymmetry algebra on the lattice does not
close, to ensure the supersymmetry of the Hamiltonian. Moreover,
due to the contribution of doubler states, the non-invariant terms
do not just fail to vanish in the continuum limit but, on the contrary,
they even diverge!

We have considered a simplified toy model related to the bit string.
However, this model catches, besides technical details, the crucial
properties of the string bit model under study. The result,
also, is not an unexpected one. Firstly, because the Poincar\'{e}
algebra which is important part of the supersymmetry algebra is
gravely affected by the discretization. (In our case it is, in fact,
reduced to continuous shifts in time and discrete one in the spatial
direction, while rotations are completely lost). It would be
at least strange if it were otherwise, because the string bit
model can be (consistently if there was no doubling)
formulated in any dimension and any background what comes in
contradiction with the fact that consistent superstring theories
can exist only in very special spaces and
backgrounds.\footnote{Strictly speaking, this is related not only
to the fermion doubling problem but also to violations of conformal
symmetry on the lattice.}
\section{Bit string quantization (\`{a} la Metsaev)}

Let us solve the equations of motion, following
\cite{Metsaev:2001bj}. This will allow us to quantize the bit
string and understand the phenomenon of doubling.

The equations of motion arising from the action \eqref{action}
read
\begin{equation}\label{bos-eom}
  -\ddot{x}^i_n+\frac{1}{2a^2}(x_{\gamma(n)}-2x_n+
  x_{\gamma^{-1}(n)})-x^i_n=0,
\end{equation}
for the bosonic part, and
\begin{subequations}\label{ferm-eom}
\begin{align}
  &\dot{\theta}_n+\frac{1}{2a}(\theta_{\gamma(n)}-\theta_{\gamma^{-1}(n)})+\Pi\q=0,\\
  &\dot{\q}_n-\frac{1}{2a}(\q_{\gamma(n)}-\q_{\gamma^{-1}(n)})-\Pi\theta=0,
\end{align}
\end{subequations}
for fermions.
Once again, let us limit ourselves to the one-string sector and fix the
class $[\gamma_1]$ by the standard choice: $\gamma(n)=n+1\mod J$.
As we discussed earlier, the solution corresponding to an
arbitrary element of the class is obtained by permutation of
labels in the ``standard'' solution.

The solution to the equations of motion is obtained in a way
analogous to that of \cite{Metsaev:2001bj}, except the discrete
Fourier transform \eqref{fourier} is used. In particular, the
bosonic part of the solution looks as follows:
\begin{equation}\label{sol-b}
  x^i_n(\tau)=X^i\cos\tau+P^i\sin\tau+\sum_{l=\pm 1,\dots,\pm [J/2]}
  \frac{1}{\omega_l}(\alpha_l^{1i}\hat{\varphi}^1_{l;n}(\tau)+
  \alpha_l^{2i}\hat{\varphi}^2_{l;n}(\tau)),
\end{equation}
where $\alpha_l^{ai}$ are string mode operators while
$\varphi^a_{l;n}(\tau)$ are respective modes of the string
\begin{subequations}\label{hatted}
\begin{align}
  &\hat{\varphi}^1_{l;n}(\tau)=\exp(-\ii(\hat{\omega}_l\tau-2\pi ln/J))\\
  &\hat{\varphi}^2_{l;n}(\tau)=\exp(-\ii(\hat{\omega}_l\tau+2\pi ln/J)),
\end{align}
\end{subequations}
and
\begin{equation}
  \hat{\omega}_l=\sgn l\sqrt{\hat{k}_l^2+1},\qquad
  \hat{k}^2_l=\frac{2}{a^2}\left(1-\cos\frac{2\pi l}{J}\right).
\end{equation}
Once again, it is not difficult to see that, as $J\to\infty$,
$a=1/J\to 0$, one recovers the solution of \cite{Metsaev:2001bj}.

Let us turn now to the fermionic sector. The solution in this
sector reads
\begin{subequations}\label{sol-f}
\begin{align}
  &\theta_n(\tau)=\cos\tau \Theta +\sin\tau \Pi
  \tilde{\Theta}+\sum_{l}c_l
  (\check{\varphi}^1_{n;l}(\tau)\theta^1_l+\ii
  (\check{\omega}_l-\check{k}_l)\check{\varphi}^2_{n;l}(\tau)\Pi\theta_l^2)\\
  &\q_n(\tau)=\cos\tau \tilde{\Theta} +\sin\tau \Pi
  \Theta+\sum_{l}c_l
  (\check{\varphi}^2_{n;l}(\tau)\theta^2_l-\ii
  (\check{\omega}_l-\check{k}_l)\check{\varphi}^1_{n;l}(\tau)\Pi\theta_l^1),
\end{align}
\end{subequations}
where, as in the bosonic case, the sum is performed over $l=\pm
1,\dots,\pm [J/2]$, and the fermionic modes
$\check{\varphi}^a_{n;l}(\tau)$ are given by the same expressions
\eqref{hatted}, except that the hatted $\hat{\omega}_l$ and
$\hat{k}_l$ are replaced by the ``checked'' ones
$\check{\omega}_l$, $\check{k}_l$, given by
\begin{equation}
  \check{\omega}_l=\sgn l\sqrt{\check{k}_l^2+1},\qquad
  \check{k}_l=\frac{2}{a}\left(\sin\frac{2\pi l}{J}\right).
\end{equation}
The peculiarity of the fermionic solution \eqref{sol-f} is that,
owing to the presence of a sin$^2$ factor (instead of cos, as in
the bosonic case), very high fermionic modes $l\sim J/2$ possess
the same energy as the modes in the region $l\ll J$. In fact, the
modes of the same energy come in pairs $(l,J/2-l)$, in total
accord with the discussion of the previous section.
The canonically quantized model is obtained by replacing the
Poisson brackets of the oscillator modes generators $\alpha^a_l$
and $\theta^a_l$ (where $a=1,2$) with the commutation relations
\cite{Metsaev:2001bj,Metsaev:2002re}
\begin{equation}\label{quant-b}
  [P^i,X^j]=-\ii\delta^{ij},\quad [\alpha^{ai}_l,\alpha^{bj}_m]=
  \frac{1}{2}\hat{\omega}\delta^{ab}\delta^{ij}\delta_{m+n,0},
\end{equation}
for bosonic modes, and
\begin{equation}
  \{\theta^{a\alpha}_l,\theta^{b\beta}_m\}=
  -\frac{1}{4}\delta^{ab}\delta^{\alpha\beta}\delta_{m+n,0},
\end{equation}
for fermionic ones.

A note is in order. The solution we found in this section
corresponds to a particular choice of the cyclic permutation
$\gamma(n)$. As proposed in \cite{Verlinde:2002ig}, the physical
states of the string bit model are those symmetrized with respect
to conjugations of $\gamma$, $h^{-1}\gamma h$, or averaged over
the conjugacy class of $\gamma$. As we noted earlier, going to a
different $\gamma$, in the same conjugacy class, is equivalent to
a permutation of the labels $n\to h(n)$ \cite{Scott:GT}.
Therefore, a solution with a different $\gamma'=h^{-1}\gamma h$ is
still given by Eqs. \eqref{sol-b} and \eqref{sol-f}, where now the
functions $\varphi_{n;l}$ are replaced by $\varphi_{h(n);l}$.
Then, a physical state with $B$ bosonic and $F$ fermionic modes
symmetrized over the permutations generically looks as follows:
\begin{equation}
  \frac{1}{J!}\sum_{h\in S_J}\alpha^{h^{-1}\gamma h}_{l_1}\dots
  \alpha^{h^{-1}\gamma h}_{l_B}\theta^{h^{-1}\gamma h}_{l_1}\dots
  \theta^{h^{-1}\gamma h}_{l_F}\ket{0},
\end{equation}
where the labels correspond to raising operators.
The ground state is unique and invariant with respect to the
permutation group $S_J$, so we do not have to twist the vacuum.


\section{BMN correspondence}

So far, we observed that the bit string model contains a number of
problems like fermion doubling and supersymmetry violation. On the
other hand, the bit string model is equivalent to BMN sector of
the super--Yang--Mills model at any finite $J$. This equivalence
would imply that the fermionic subsector of the BMN operators is badly
defined, at finite $J$. (Since there is no definition for the BMN
correspondence at $J=\infty$, this would signal a self-consistency
problem in the BMN correspondence.) Hence, in this section we
proceed to the analysis of the implications of fermion doubling at
the level of the BMN operators.

The BMN correspondence \cite{Berenstein:2002jq} relates a class of
operators in $\mathit{N}=4$ super--Yang--Mills model, which have a
large $R$-charge ($J\to\infty$), to states in the closed
superstring on the pp-wave background.
The string ``semantics'' of the BMN language is as follows. The
light-cone superstring vacuum in the BMN language is given by the
operator
\begin{equation}\label{BMN-ground}
  \frac{1}{\sqrt{J}N^{J/2}}\tr[Z^J] \leftrightarrow \ket{0,p_{+}},
\end{equation}
where $Z$ is the complex scalar component,
$Z=(\phi^5+\ii\phi^6)/\sqrt{2}$.

The excited string states correspond to the insertion of
``impurities'' under the trace \eqref{BMN-ground}, according to
the following rule:
\begin{align}
  &D_\mu Z \leftrightarrow \alpha^{\dag\mu},\qquad &\mu=1,\dots,4\\
  &\phi^{j-4} \leftrightarrow \alpha^{\dag i},\qquad &i=5,\dots,8\\
  &\chi^a_{J=\frac{1}{2}}\leftrightarrow \theta^{\dag \alpha},\qquad
  &\alpha=1,\dots,8,
\end{align}
where $\alpha^\dag$ and $\theta^{\dag}$ are, respectively, bosonic
and fermionic standard oscillator raising operators. Also, in
order to get
nonzero string modes, the insertions should be accompanied by a
factor $\e^{\frac{2\pi\ii kn}{J}}$, where $k$ is the position of
the insertion in the row of $Z$'s. Hence, e.g. a double fermionic
insertion corresponds to
\begin{equation}
  \frac{1}{\sqrt{J}N^{J/2+1}}\sum_{k=0}^{J-1}\tr
  [\chi^\alpha_{J=\frac{1}{2}}Z^k\chi^\beta_{J=\frac{1}{2}}Z^{J-l}]
  \e^{\frac{2\pi \ii kl}{J}}\leftrightarrow
  \theta^{\dag\alpha}_l\theta^{\dag\beta}_{-l}
  \ket{0,p_+}.
\end{equation}

Thus, the BMN correspondence (\emph{language}) is formulated in
terms of \emph{words} or strings of products of $Z$'s, with
insertions of impurities and operators on the space of allowed
words. The main operators acting on words are the position $X$ and
the shift $P$. $X$ gives the position $j$ (up to a cyclic
permutation) of an insertion in the chain of $Z$'s, while $P$
performs a permutation of the impurity in the $j$-th position to
the $(j+1)$-th one.

One can define the scalar product of words $\Psi\sim\dots Z\phi
Z\dots\psi\dots$, which is given by
\begin{equation}
  (\Psi,\Psi')=\langle \overline{\Psi}\Psi'\rangle_{N,J\to\infty}.
\end{equation}
(The necessary properties required for this to be a scalar product
follow from the planar properties of the correlator in the large
$N$ limit, \cite{Berenstein:2002jq}.)
As it can be seen, the shift operator $P$ is not self-adjoint,
with respect to the BMN scalar product, and there is an adjoint
operator $P^+$, which corresponds to the backward shift
\begin{equation}
  P^+:j\to j-1.
\end{equation}

Since the bosonic interaction comes through the term $\sim
g^2_{YM}\tr[Z,\phi][\bar{Z},\phi]$, this produces in the bosonic
part of the effective Hamiltonian a term proportional to $P^+P$
(due to the cyclic property of the trace this is the same as
$PP^+$). On the other hand, the fermionic interactions in
super--Yang--Mills theory are linear in the shifts
\begin{equation}
  \sim (\chi\Gamma_Z[Z,\chi]+\chi\Gamma_{\bar{Z}}[\bar{Z},\chi]).
\end{equation}
This leads to a contribution proportional to the symmetric part of
the shift operator $\sim 1/2(P+P^+)$ in the fermionic part of the
effective Hamiltonian. As we observed above, when analyzing the
bit string model, this leads to the fermionic spectrum doubling.
In terms of the shift operators, this is explained by the
existence of such zero modes of $1/2(P+P^+)$, which correspond to
highly oscillating modes on the lattice string.

\section{Discussion}
In this paper we addressed the problem of finite $N$ effects in
the BMN correspondence. For finite $N$ and $J$, the set of BMN
operators maps into the Hilbert space of $J$ string bits. As we
have shown above, the fermionic spectrum of the string bit model
is doubled. An immediate effect of doubling is the failure to get
a supersymmetric limit, as $J\to\infty$.

We considered a free theory and, on this level, one can explicitly
separate the contribution of the doubler states, in order to get
the correct spectrum of IIB string, as $J$ and $N$ go to infinity.
We \emph{believe} that this can also be done on the tree level of
interacting closed superstrings. However, as the experience of the
lattice shows, in the case of bit loops the doubling states mix
with the correct modes.

In spite of above problems, the study of supersymmetric models on
the lattice have achieved, during the last several years, a
considerable progress (see
\cite{Montvay:1998ak,Montvay:2001aj,Kaplan:2002zs} for a
review).\footnote{While this work was in progress a paper
\cite{Catterall:2003wd} studying topological models on the lattice
appeared.} One can hope to apply the technique developed in this
approach to string bits too. This is accompanied, however, by the
fact that, beyond the typical lattice problem with fermion doubling,
there are specifical string problems, related to conformal
invariance violation by the string discretization. One can also
expect the duality symmetries to be violated too.

Returning to the BMN correspondence, one can see that there is a
class of unwanted fermionic states, given by the fermionic
doublers, which survive in the (formal) BMN limit. In fact, the
BMN sector is known to contain some ``extra'' states which are
conjectured to decouple because of large masses acquired due to
the interactions \cite{Berenstein:2002jq}. On the other hand, the
above arguments about decoupling, used in \cite{Berenstein:2002jq}
would hardly apply to fermion doublers, since they propagate and
interact exactly in the same way, as the genuine fermionic modes.

The above fact can also signal the presence of the same problem in
the fermionic spectrum of the AdS/CFT correspondence, when one
tries to obtain such correspondence starting from large but finite
values of $N$. In order to be able to say something more precise, one has
to study this topic too. We hope to do this in the future.
Perhaps one can avoid the problems, working directly in the model
with $N=\infty$, which is the super--Yang--Mills model in
noncommutative space. However, in this case, a procedure allowing
to get rid of non-planar contributions must be devised and
implemented.

\subsection*{Acknowledgements}
Useful discussions with Francisco Morales, Jan Plefka and
Fabrizio Palumbo are acknowledged.

This work was partially supported by a RBRF grant, the European
Community's Human Potential Programme under the contract
HPRN-CT-2000-00131 Quantum Spacetime, the INTAS-00-0254 \&
INTAS-00-0262 grants, the NATO Collaborative Linkage Grant
PST.CLG.979389 and the Iniziativa Specifica MI12 of the INFN
Commissione Nazionale IV.


\end{document}